\documentclass[apj]{emulateapj}

\usepackage{epsfig}
\usepackage{natbib}

\begin{document}
\newcommand{\kms}{km~s$^{-1}$}
\newcommand{\Msun}{M_{\odot}}
\newcommand{\Lsun}{L_{\odot}}
\newcommand{\ML}{M_{\odot}/L_{\odot}}
\newcommand{\etal}{{et al.}\ }
\newcommand{\hhh}{h_{100}}
\newcommand{\hsq}{h_{100}^{-2}}
\newcommand{\tn}{\tablenotemark}
\newcommand{\mdot}{\dot{M}}
\newcommand{\p}{^\prime}
\newcommand{\kmsMpc}{km~s$^{-1}$~Mpc$^{-1}$}

\title{Cosmicflows-2: SNIa Calibration and H$_0$}

\author{H\'el\`ene M. Courtois}
\affil{Institute for Astronomy, University of Hawaii, 2680 Woodlawn Drive, Honolulu, HI 96822, USA\
University of Lyon; UCB Lyon 1/CNRS/IN2P3/INSU; Lyon, France}

\and
\author{R. Brent Tully,}
\affil{Institute for Astronomy, University of Hawaii, 2680 Woodlawn Drive, Honolulu, HI 96822, USA}

\begin{abstract}
The construction of the Cosmicflows-2 compendium of distances involves the merging of distance measures contributed by the following methods: (Cepheid) Period-Luminosity, Tip of the Red Giant Branch (TRGB), Surface Brightness Fluctuation (SBF), Luminosity-Linewidth (TF), Fundamental Plane (FP), and Type Ia supernova (SNIa).  The method involving SNIa is at the top of an interconnected ladder, providing accurate distances to well beyond the expected range of distortions to Hubble flow from peculiar motions.  In this paper, the SNIa scale is anchored by 36 TF spirals with Cepheid or TRGB distances, 56 SNIa hosts with TF distances, and 61 groups or clusters hosting SNIa with Cepheid, SBF, TF, or FP distances.  With the SNIa scale zero point set, a value of the Hubble Constant is evaluated over a range of redshifts $0.03 < z < 0.5$, assuming a cosmological model with $\Omega_m = 0.27$ and $\Omega_{\Lambda} = 0.73$.  The value determined for the Hubble Constant is H$_0 = 75.9 \pm 3.8$~\kmsMpc. 
\end{abstract}

Subject headings: distance scale; galaxies: distances and redshifts

\section{Introduction}

The light curves of Type Ia supernovae (SNIa) can be interpreted to give remarkably accurate distances to galaxies out to redshifts of order unity   \citep{1998AJ....116.1009R, 1999ApJ...517..565P}.  The useful reach of SNIa extends well beyond the domain of suspected velocity perturbations to the cosmic expansion so provides the best means currently available for the determination of the Hubble Constant.  However the zero point scale of the SNIa distance measurements must be established through comparisons with alternative methods and the number of local SNIa that have been observed in sufficient detail is still small.  \citet{2009ApJS..183..109R, 2011ApJ...730..119R} have undertaken the most substantial effort to date to provide the required calibration.  The current study was motivated by the opportunity to significantly enlarge the calibration sample.

This project is coupled with a program that combines new observations and archival retrieval in the construction of a compendium of galaxy distances that we are calling Cosmicflows-2.  A core element of the compendium is distances derived from the correlation between galaxy luminosities and rotation rates  
\citep{1977A&A....54..661T}.  A new calibration of this so-called TF method has recently been published \citep{2012arXiv1202.3191T}.  We have been engaged in optical photometry and radio HI observations of several samples \citep{2011MNRAS.tmp..796C, 2011MNRAS.414.2005C}, among them a sample of galaxies within 10,000 km/s that have hosted well observed SNIa.

Distances from the TF method determined within our program provide an anchor but the assembly of the Cosmicflows-2 database provides more extensive resources.   A very important component from the literature derives from the same methodology but uses independent analysis procedures; the large compilation of distances  identified by the project name `SFI++' \citep{2007ApJS..172..599S}.  Another important component is based on the Fundamental Plane methodology and here we draw on results from three groups identified by the team names `SMAC', `EFAR', and `ENEAR' \citep{2001MNRAS.327..265H, 2001MNRAS.321..277C, 2002AJ....123.2159B}.

Distance measures based on the TF and FP methods are plentiful but individually uncertain.  The situation is improved by averaging estimates for all galaxies in a common group.  In several groups or clusters, more than one SNIa has been usefully recorded.  There are instances where we do not have an alternative distance estimate to the host galaxy of a SNIa event but we do have an alternative distance estimate to the host group.

If we are going to establish a zero-point scale for the SNIa methodology then we want to have a coherent assembly of supernovae from nearby to far away.  We draw upon the UNION2 compilation \citep{2010ApJ...716..712A}.  Four local compilations based on alternative analysis variations provide checks \citep{2006ApJ...647..501P, 2007ApJ...659..122J, 2009ApJ...700.1097H, 2010AJ....139..120F}.

We begin with an assembly of a catalog of SNIa with a common set of relative distance measures.  We then give a description of our observational program and provide a preliminary calibration of the SNIa scale with our data alone.  Then there is a section that describes the integration of literature contributions and the correlation of galaxies with groups.  In the following section we are able to generate our best calibration of the SNIa scale.  

\section{A SNIa Sample}

If there is a scale that achieves a `fair sample' representation of the mean properties of the universe, distances independent of redshift on that scale are surely - and at this time uniquely - probed by SNIa.  We begin this study by accumulating a large body of SNIa distance estimates on a consistent relative scale, ranging from the nearest known SNIa to those that probe cosmological curvature at $z \sim 1$.  Our goal is to refine the absolute scale through overlaps with other distance measurement methodologies so we give particular attention to achieving good coverage locally.

We have no expertise with SNIa technology so we accept measurements from the literature and limit our attention to assuring that information from alternative sources are on a consistent scale and, through inter-comparisons, weed out a very small number of inconsistent measures.  The SNIa compilation UNION2 \citep{2010ApJ...716..712A} provides a useful reference standard because it is very large and spans the full range of observed redshifts.  The distance moduli from this reference are nominally consistent with a value of the Hubble Constant H$_0 = 100$~\kmsMpc\ and the cosmological mass-energy density parameters $\Omega_m =  0.27$ and $\Omega_{\Lambda} = 0.73$.  We have confirmed that these mass-energy density parameters give a good fit decoupled from modest modifications to the specification of H$_0$.

There have been a significant number of local SNIa events that are not included in the UNION2 compilation but have received attention by others.  We give consideration to four other sources   \citep{2006ApJ...647..501P, 2007ApJ...659..122J,  2009ApJ...700.1097H, 2010AJ....139..120F} for SNIa within $z = 0.1$.  There are large overlaps that permit each of the alternative sources to be transformed to the UNION2 scale with excellent precision.  Variations on the most popular supernova light curve fitters are represented.  In the case of Hicken et al. four fitting procedures are compared and we take distance measures from this reference based on the SALT2 analysis where available (appreciating that we are concerned with SNIa at low $z$, mitigating issues that arise at high $z$) and supplement with a few additional cases from the MLCS17 analysis.   The data that we have gathered on nearby SNIa from the five sources are made available on-line in a way that can be easily compared at EDD, the Extragalactic Distance Database\footnote{http://edd.ifa.hawaii.edu}. 

\begin{figure}[!]
\centering
\includegraphics[scale=0.43]{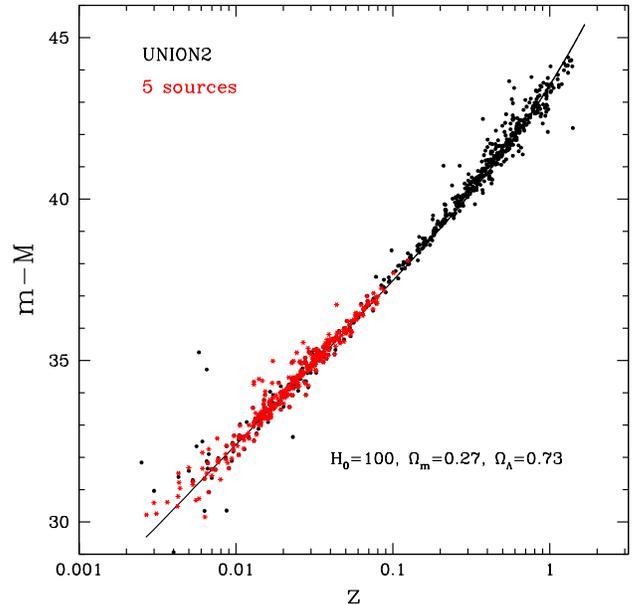}
\caption{Hubble diagram with the UNION2 sample (black circles) shown over its full redshift-distance modulus range and the average of 5 sources (red stars) shown over the range extending to $z=0.1$.  The solid line gives the UNION2 preferred cosmological fit.}
\label{hub_union}
\end{figure}

Table~\ref{tbl:snscales} records the details of correlations between UNION2 and the four additional literature sources.  At worst, rms scatter is $\sim 0.2$ mag and at best it is 0.05 mag.  Of course, different observers are looking at the same supernovae and most often the same light curve information but with separate analysis procedures.  The good news for our purposes  is that we can reliably combine the alternative sources into a single consistent catalog of distances.  Figure~\ref{hub_union} shows the Hubble diagram for the full sample with the UNION2 compilation extending to $z \sim 1$ and the combination of five sources contributing 318 SNIa within the domain $z < 0.1$.

\begin{deluxetable}{lrcccc}
\tablenum{1}
\tablecaption{Comparisons to Bring Alternate SNIa Sources to UNION2 Scale}
\label{tbl:snscales}
\tablewidth{0in}
\tablehead{\colhead{Source} & \colhead{\#} & \colhead{RMS} & \colhead{Offset} & \colhead{St.Dv.} & \colhead{Reject}}
\startdata
Prieto et al. 2006    &  79 & 0.16 & 0.712 & 0.018 & 1 \\
Jha et al. 2007        &  91 & 0.17 & 0.933 & 0.018 & 4 \\
Hicken et al. 2009  & 160 & 0.07 & 0.895 & 0.006 &   \\
Folatelli et al. 2010 &  12 & 0.05 & 0.637 & 0.013 & 1 \\
\enddata
\tablecomments{
Offset required to take source modulus $\mu_{source}$ to UNION2 scale, r.m.s. scatter in source-UNION2 comparison, and standard deviation are in units of magnitude.
$\mu_{union2} = \mu_{source} - {\rm offset}$}
\end{deluxetable}

\section{Preliminary SNIa Zero Point Calibration}

SNIa occur in all types of galaxies including normal spirals that are good targets for distance measurements based on the correlation between luminosity and rotation rate, the TF method.  We initiated a program to collect $I$ band photometry \citep{2011MNRAS.tmp..796C} and radio HI profile information \citep{2009AJ....138.1938C, 2011MNRAS.414.2005C} for spiral galaxies that have hosted SNIa drawn from the lists of \citet{2003ApJ...594....1T} and \citet{2007ApJ...659..122J}.  Some more recent supernovae have occurred in galaxies that by good fortune are found in our expanding compendium of photometry and HI spectroscopy (see EDD).  Concurrently, we have completed a re-calibration of the TF correlation (Tully \& Courtois 2012) that involves a revised procedure for handling digital HI profiles and is based on a 13 cluster slope template and a 36 galaxy zero point calibration specified by Cepheid Period-Luminosity \citep{2001ApJ...553...47F} and Tip of the Red Giant Branch \citep{2007ApJ...661..815R} distance measures.

\begin{figure}[!]
\centering
\includegraphics[scale=0.43]{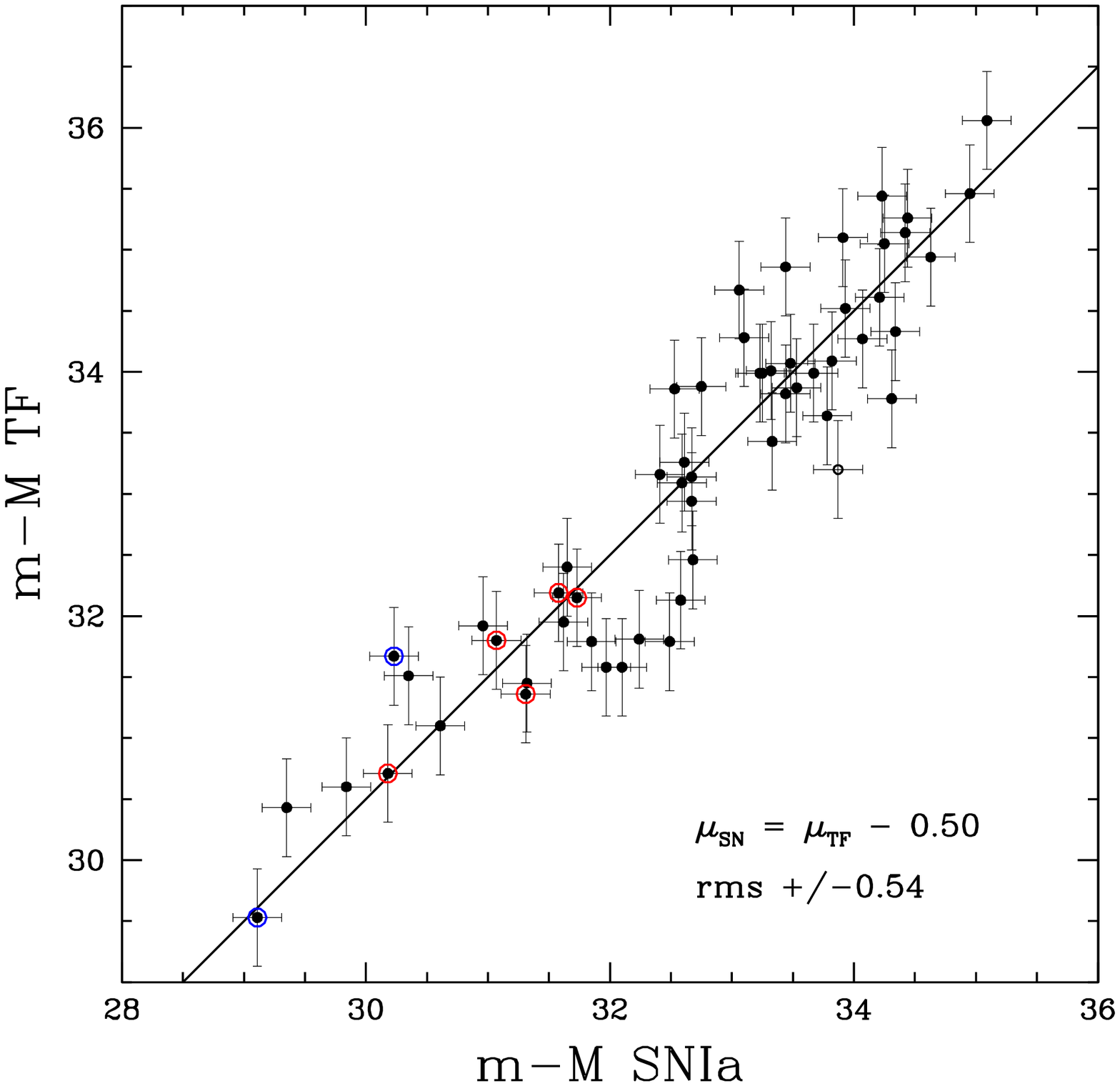}
\includegraphics[scale=0.43]{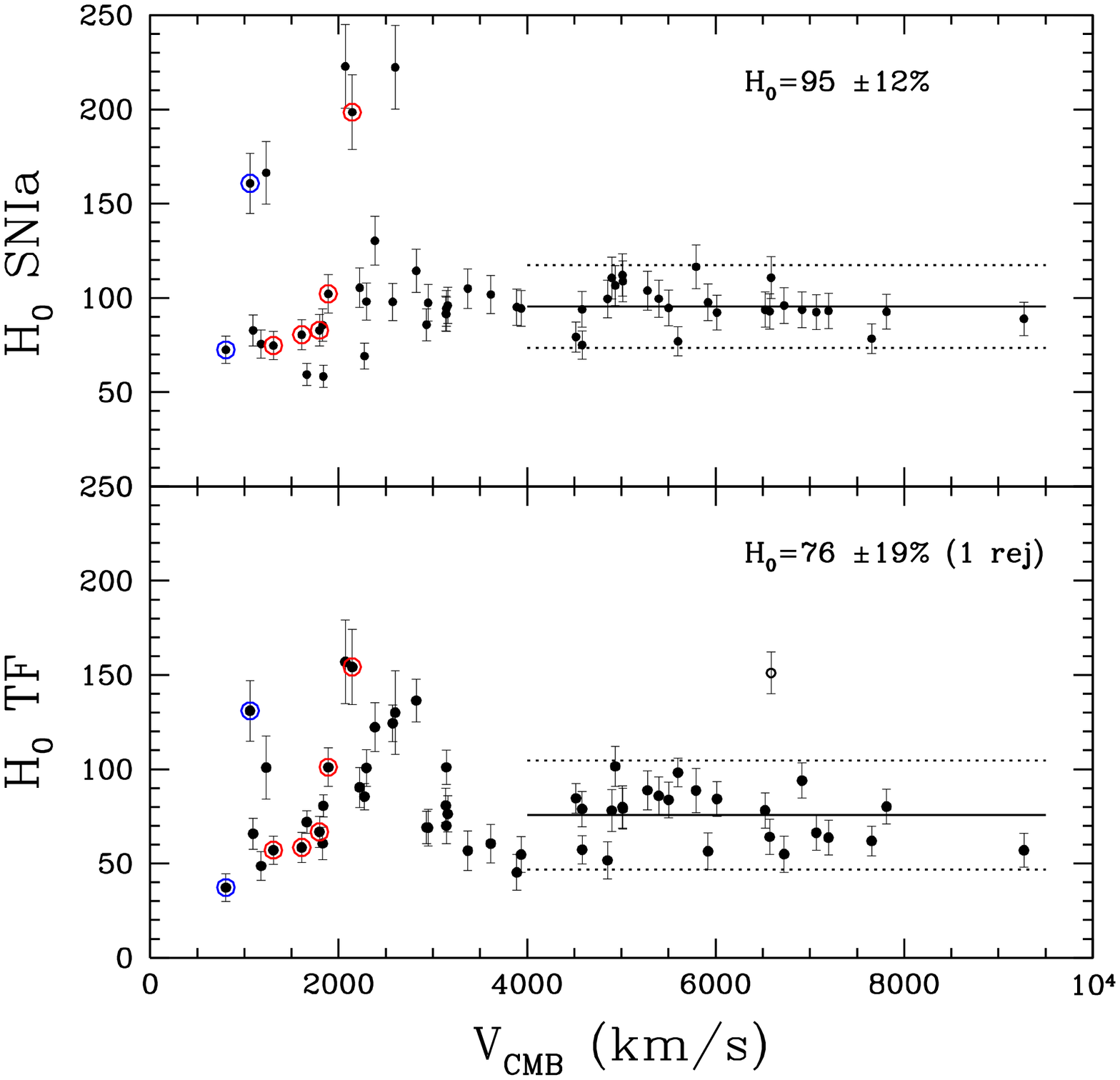}
\caption{{\it Top:} Correlation between SNIa and TF distance moduli with an assumed slope of unity.  The TF scale is established by 36 galaxies with Cepheid or TRGB distances and the SNIa scale is based on systemic velocities and an assumed H$_0 = 100$.  From the displacement of 0.50 mag, a preliminary rescaling of SNIa distances suggests H$_0 = 79$~\kmsMpc. The 5 cases identified by red circles are among the 8 galaxies in the Cepheid calibration sample of \citet{2011ApJ...730..119R} and the two cases identified by blue circles correspond to additional galaxies with Cepheid distance measures.  Error bars assume uncertainties of 0.4 mag in TF moduli and 0.2 mag in SNIa moduli.
{\it Bottom:} Hubble parameter (velocity/distance) for the same sample of 36 galaxies, with distances from SNIa used in the upper panel and TF distances used in the lower panel.  The mean values of the Hubble parameter for $V_{CMB}>4000$~\kms\ of 95 and 76 \kmsMpc\ respectively and the $2\sigma$ scatter range are indicated by horizontal lines.  One discordant point given an open symbol is rejected from the fits.}
\label{sn_tf}
\end{figure}

Presently we have TF distances for 56 galaxies that have hosted well observed SNIa.  These galaxies are identified and information about the distance estimates are given in Table~\ref{tbl:sn_tf}.  A comparison of TF and SNIa distance is shown in Figure~\ref{sn_tf}.

We are surprised by two aspects of this comparison.  First, the scatter, at $\pm0.54$ mag rms, is larger than anticipated by the fiducial expectation of TF scatter $\sim 0.4$ and SNIa scatter $\sim 0.2$.  What is seen in the lower panel of Fig.~\ref{sn_tf} does not resolve the issue.  Restricting attention to velocities larger than 4000~\kms\ to minimize influence of peculiar velocities, the scatter found in the Hubble parameter $-$ velocity/distance $-$ are consistent with expectations from the SNIa and TF measures respectively.  A scatter of $\sim0.45$ mag would be anticipated in the upper panel.
Second, taken from this SNIa sample alone and accepting the nominal UNION2 zero point as a valid fit, it would be inferred that H$_0 = 79$~\kmsMpc\ which is higher than most modern estimates.  However the UNION2 zero point set at H$_0=100$ is nominal and we will see that it is best fit by a slightly lower value.

\section{Alternative Distances and Group Averaging}

The comparison against the SNIa scale in the last section involved only our own measurements but the comparison can be broadened considerably by involving distance estimates from other sources.  In addition, the linkage of SNIa events with galaxy groups allows, first, for an averaging of distance and velocity parameters over multiple measures and, second, for an extension of comparisons beyond the specific galaxy that hosted the SNIa.

In a few very nearby cases there are high precision distance determinations.  Our fundamental zero-point is established in concordance with the HST Key Project scale assuming a modulus of 18.5 for the Large Magellanic Cloud  \citep{2001ApJ...553...47F}.  There are direct Cepheid measurements for 12 SNIa hosts, including 8 used by \citet{2011ApJ...730..119R} in their determination of the SNIa distance scale zero point. These and other Cepheid observations provide distances to the Virgo, Fornax, and Centaurus clusters.  The Surface Brightness Fluctuation (SBF) method provides alternative precision distances on the same scale to the Virgo and Fornax clusters  \citep{2009ApJ...694..556B} and to a dozen smaller groups or clusters \citep{2001ApJ...546..681T, 2010ApJ...724..657B}.

Next, we consider measures that are individually less precise but are numerous and effective over a wide range of redshifts.  One particularly important source to consider uses the same TF methodology discussed in the previous section: the extended field spiral sample with $I$ band photometry (SFI++) discussed by \citet{2007ApJS..172..599S}, a collection and extension of results from the Cornell group \citep{1997AJ....113...53G,  1999AJ....118.1489D, 2006ApJ...653..861M, 2007AJ....134..334C}.  Then extending to a separate methodology, we give attention to three sources that independently work with the Fundamental Plane (FP) relationship.  The three contributing teams describe themselves with the acronyms EFAR  \citep{2001MNRAS.321..277C}, SMAC \citep{2001MNRAS.327..265H}, and ENEAR \citep{2002AJ....123.2990B, 2002AJ....123.2159B}.

\begin{figure}[b]
\centering
\includegraphics[scale=0.43]{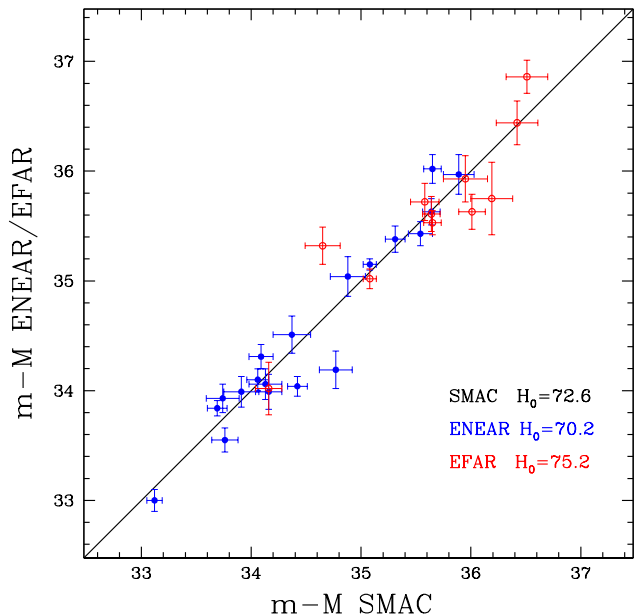}
\caption{Inter-comparison of three sources of distance measurements with the Fundamental Plane technique for samples drawn from clusters.  SMAC distance moduli with a zero point nominally consistent with H$_0 = 72.6$~\kmsMpc\ provide a reference on the horizontal axis.  On the vertical axis, ENEAR comparisons are shown by filled blue symbols and EFAR comparisons are shown by open red symbols.  Best fits required shifts from published moduli to nominal values of H$_0 = 70.2$ for ENEAR and H$_0 = 75.2$ for EFAR.}
\label{3fp}
\end{figure}

Attention is first given to the integration of the three FP samples. There is essentially no overlap between EFAR and ENEAR samples but both have considerable overlap with SMAC.  \citet{2001MNRAS.327.1004B} established a zero point link between SBF and SMAC measurements.  For these reasons we begin by using SMAC as a basis of comparison between FP sources.  The comparisons binned to clusters are shown graphically in Figure~\ref{3fp} and details are recorded in Table~\ref{tbl:compare}.  The value of the Hubble parameter indicated in the case of SMAC comes from the SBF linkage and the values of the Hubble parameter indicated in the cases of EFAR and ENEAR come from shifts from the EFAR and ENEAR nominal zero points resulting from the least squares fits to the overlaps with SMAC.  Each of the individual correlations is consistent with a slope of unity.

The comparisons in Figure~\ref{3fp} are averaged measures over multiple galaxies in clusters, mostly from the revised Abell catalog \citep{1989ApJS...70....1A}.  Similarly with SFI++, our interest in this paper is with averaged measures within clusters.  Within this latter data set, we accept the `In' memberships identified by the SFI++ collaboration \citep{2006ApJ...653..861M, 2007ApJS..172..599S}.  We then seek to minimize the zero point differences in the three-way overlap of distances to clusters in the Cosmicflows-2 calibration paper (CF2) of Tully \& Courtois (2012), the SFI++ clusters, and clusters in the composite of the three Fundamental Plane sources (FP).

\begin{figure}[!]
\centering
\includegraphics[scale=0.43]{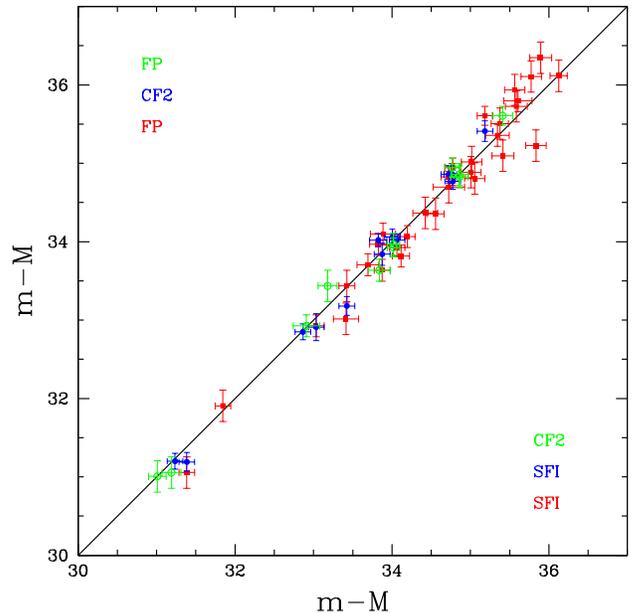}
\caption{Three way comparison between CF2, SF++, and FP sources of distances to clusters.  The green open circles compare CF2 on the horizontal axis with FP on the vertical axis.  The blue filled circles compare SFI++ (horizontal) and CF2 (vertical).  The red filled squares compare SFI++ (horizontal) and FP (vertical).  Zero points are set to the scale established by CF2.}
\label{3meth}
\end{figure}

With each comparison of distances, it was first confirmed that an unrestricted slope is statistically compatible with a slope of unity.  Then, with a slope of unity assumed, zero point offsets were determined between the pairwise combinations of the triad CF2, SFI++, and FP.  Results are recorded in Table~\ref{tbl:compare}.  The final zero point is set by CF2.  Operationally, we first merged CF2 and SFI++ because these two are based on the same methodology and the rms scatter and standard deviation are the smallest of the pairwise comparisons.  Then we took the average of the shifts necessary to bring FP into separate agreement with CF2 directly (shift: 0.145 mag) and to SFI++ then to CF2 (shifts: 0.077+0.045=0.122 mag).  The difference of 0.023 translates to 1\% in distance.  We take the straight average of the two shifts to bring FP to the CF2 scale because the standard deviations of the two routes are essentially equal: the smaller dispersion in the case of direct to CF2 is offset by the larger comparison sample in the case of passage by SFI++.  The three-way comparisons are shown in Figure~\ref{3meth} set to our final preferred zero point scale.

\begin{deluxetable}{lrcrc}
\tablenum{3}
\tablecaption{Source Intercomparisons}
\label{tbl:compare}
\tablewidth{0in}
\tablehead{\colhead{Sources} & \colhead{\#} & \colhead{RMS} & \colhead{Offset} & \colhead{St.Dv.}}
\startdata
ENEAR$-$SMAC & 19 & 0.22 &   0.026     & 0.051 \\
EFAR$-$SMAC    &  11 & 0.31 & $-0.051$ & 0.094 \\ 
SFI$-$CF2             &  13 & 0.14 &  0.045     & 0.039 \\
FP$-$SFI               &  32 & 0.25 &  0.077     & 0.043 \\
FP$-$CF2              &  11 & 0.14 &  0.145     & 0.043 \\ 
\enddata
\tablecomments{
$\mu_{1st}  - \mu_{2nd} = {\rm offset}$}
\end{deluxetable}

The revised SFI++ and FP scales are small departures from values advocated by the relevant authors.  In the case of SFI++, the published zero point \citep{2006ApJ...653..861M} is based on 16 calibrators in common with our 36.  There is a 1\% difference in the calibration because SFI++ includes metallicity corrections to the Cepheid magnitudes \citep{2004ApJ...608...42S}.  Excluding that minor correction to enable a direct comparison, then the value of H$_0$ that SFI++ favors is 75.  The adjustment to the CF2 scale would increase this value to 76~\kmsMpc.  In the case of the FP samples, only SMAC tried to provide a zero point calibrated scale which is why we transfered all FP values to agree with that source.  The calibration in that case was from an overlap with SBF targets that was tied back to the HST Key Project Cepheid scale.  The value of H$_0$ preferred by that source is 72.6.  The value transferred to the CF2 scale is 77~\kmsMpc.  The tentative value for H$_0$ derived in the CF2 calibration paper (Tully \& Courtois 2012) is 75~\kmsMpc.

The SFI++ and three FP samples applied to clusters provides increased reliability in the comparison we will make with SNIa distance scale but for the most part these samples, restricted as they are to clusters, make contributions only beyond $\sim 3000$~\kms.  More nearby, we can take advantage of a knowledge of the composition of groups, using an updated version of an early catalog \citep{1987ApJ...321..280T}, to permit an averaging over multiple measures and to permit comparisons if the SNIa host has no alternative distance estimate but resides in a group with a measure.  We accumulate the various distance determinations for clusters beyond 3000~\kms\ in Table~\ref{tbl:dm_gt3k} and for groups within 3000~\kms\ in Table~\ref{tbl:dm_lt3k}.  

A remaining and very important source of comparison is with the direct SNIa $-$ Cepheid linkage established by \citet{2011ApJ...730..119R}.  We accept their zero point based on the average of the maser distance to NGC 4258 \citep{1999Natur.400..539H} and anchors based alternatively on Large Magellanic Cloud or Galactic Cepheids.  The Riess et al. sample is small, only 8 galaxies, but the measures are high quality.

\begin{figure}[!]
\centering
\includegraphics[scale=0.43]{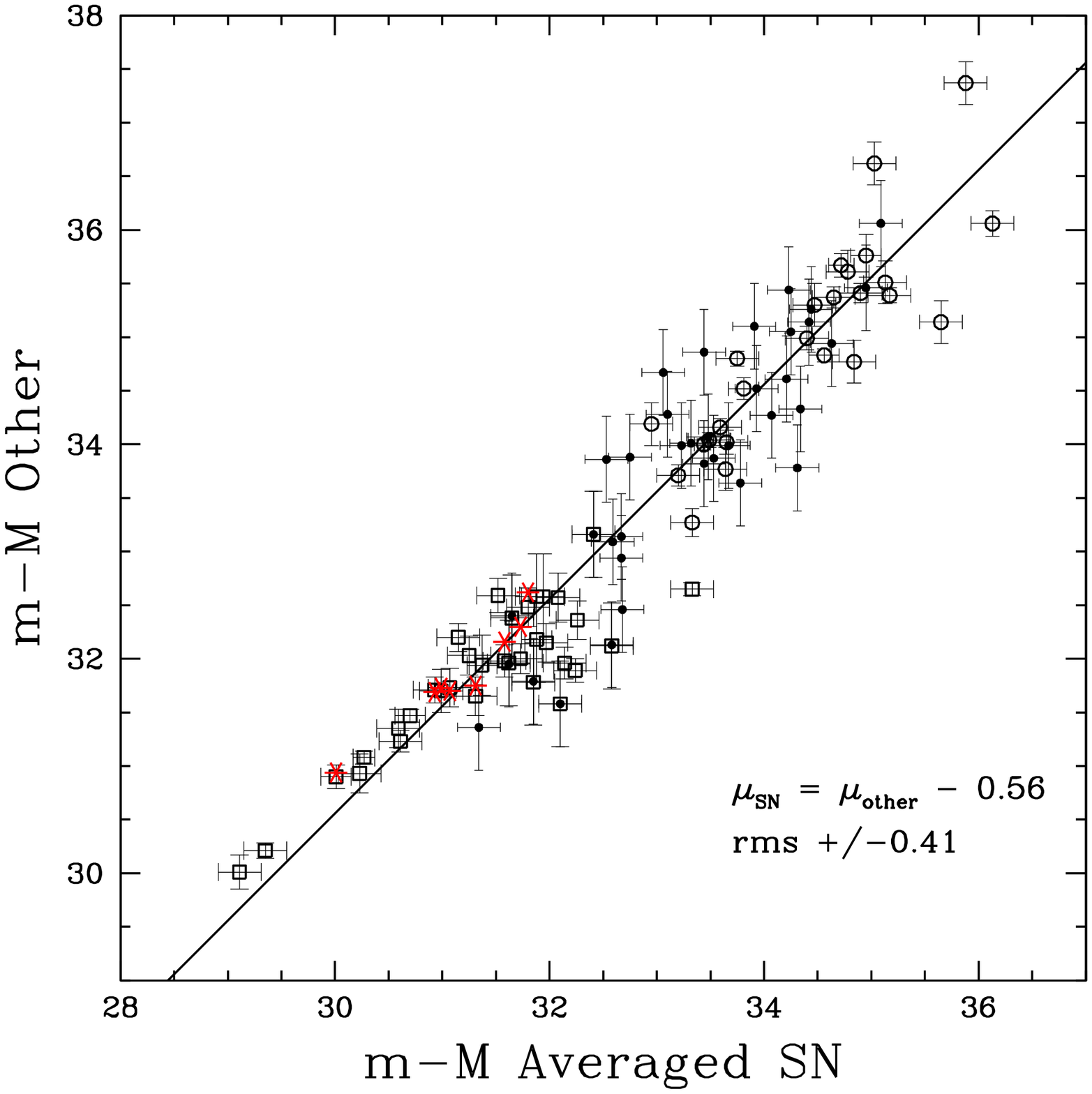}
\includegraphics[scale=0.43]{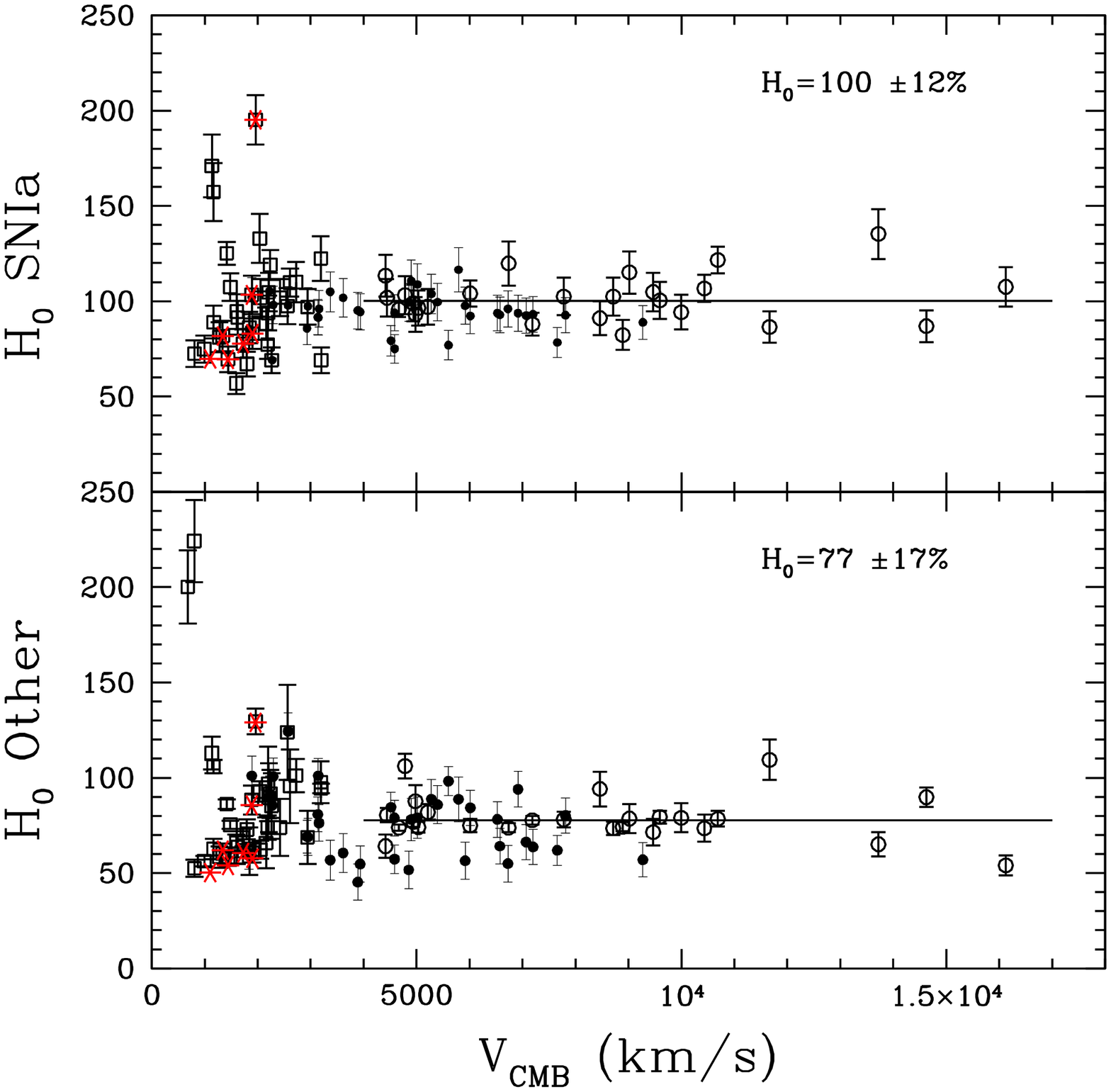}
\caption{{\it Top:} Comparison between SNIa distances and other measures after group binning.  Open circles: clusters beyond $\sim 3000$~\kms\ with distances averaged over the sources CF2, SFI++, and FP. Open squares: groups within $\sim 3000$~\kms\ with distances averaged over Cepheid, SBF, TF, and FP measures where available.  Small filled circles: individual measures drawn from Fig.~\ref{sn_tf} if group averaging is not possible. Red stars: the 8 Riess et al. Cepheid calibrators.
{\it Bottom:} Hubble parameter (velocity/distance) for the group binned samples.  SNIa distances are used in the upper panel and group or `other' distances are used in the lower panel.  The horizontal lines illustrate values of the Hubble parameter averaged over cases with $V_{CMB}>4000$~\kms\ of 100 and 77 \kmsMpc\ respectively.}
\label{dm_gp}
\end{figure}

In Figure~\ref{dm_gp} we show a comparison between distance moduli by alternative methods transformed to the CF2 scale and by SNIa on the UNION2 scale.  There is complexity to the fit because of the heterogeneous nature of the input.  We use weights $w = 1/\epsilon^2$ where $\epsilon$ are errors assigned by the following rules.  Individual SNIa, Cepheid, and HST SBF measures are given errors $\epsilon = 0.2$ in distance modulus, ground based SBF measures are given errors $\epsilon = 0.3$, while TF and FP measures are given errors $\epsilon = 0.4$.  No cluster or group is given a cumulative error less than $\epsilon = 0.1$ from a single method.  The final fit involves 36 clusters at velocities greater than 3000~\kms, 25 groups within 3000~\kms, 36 individual galaxies that hosted SNIa with TF measures, and 8 individual galaxies that have hosted SNIa with Cepheid measures.  The offset between the SNIa and `other' distance moduli is 0.557 with a scatter ${\bar \epsilon} = 1/\sqrt{{\bar w}} = 0.41$~mag.  This offset would shift the nominal SNIa scale from H$_0 = 100$ to H$_0 = 77$~\kmsMpc.  However the nominal UNION2 SNIa zero point requires attention.  The scatter in the top panel conforms to expectations implicit in the scatter in the lower panels.  The standard deviation with 69 `good' measures is 0.05 mag.

\section{The SNIa Scale and H$_0$}

\begin{figure}[!]
\centering
\includegraphics[scale=0.43]{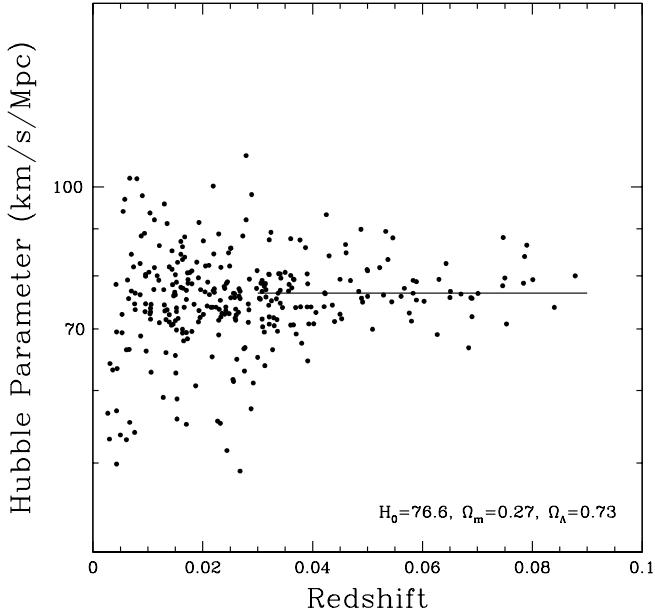}
\caption{Hubble parameter as a function of redshift for 311 SNIa in a merged catalog from 5 sources.    Solid line: best fit to 118 cases with $0.03 < z < 0.09$ assuming a cosmological model with $\Omega_m = 0.27$, $\Omega_{\Lambda} = 0.73$ and H$_0 = 76.6$. }
\label{hub_5_30k}
\end{figure}

The zero point linkage to the SNIa distance scale provides the opportunity to obtain a measure for H$_0$ in a `fair sample' domain, beyond the range of significant velocity perturbations.  In the previous section we established a bridge to the sample of SNIa with $z < 0.1$ discussed in Section 2.  This sample is referenced to the UNION2 scale but includes contributions from four additional sources.  Figure~\ref{hub_5_30k} illustrates  a variation of the Hubble diagram for this SNIa sample.  The Hubble parameter is defined by the equation
\begin{equation}
{\rm H}_0 = {{cz}\over{d}} {\{1 + {{1}\over{2}} [1 - q_0]z - {{1}\over{6}} [1 - q_0 -3 q_0^2 + j_0]z^2\}}
\end{equation}
where $d$ and $z$ are distance and redshift, we assume the jerk parameter $j_0 = 1$ and the acceleration parameter $q_0 = {{1}\over{2}} ( \Omega_m -2 \Omega_{\Lambda} ) = -0.595$. 
The zero point is revised in accordance with the fit to alternative distance measures shown in Fig.~\ref{dm_gp}.  The choice of cosmological model is not important for this relatively local sample.  The effect of using an alternate model on the preferred value of H$_0$ is at the level of 1\%.  The fit to all 311 cases in Fig.~\ref{hub_5_30k} gives H$_0 = 75.2$.  The best fit over the range $0.03 < z < 0.09$ illustrated by the flat line requires H$_0 = 76.6 \pm 0.5$~\kmsMpc.  The uncertainty is the standard deviation of the fit. The rms scatter in 118 cases is $\pm 6.9$~\kmsMpc\ (with $3 \sigma$ culling of 1 object) corresponding to 9\% in distance at a given velocity. 

\begin{figure}[!]
\centering
\includegraphics[scale=0.43]{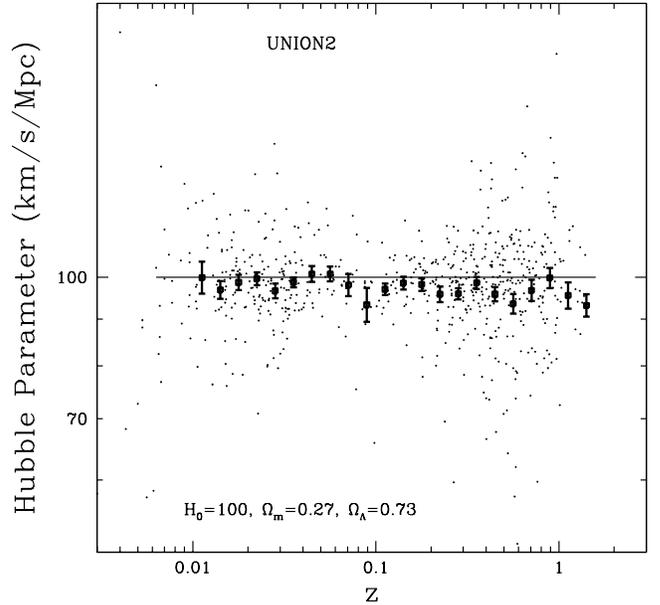}
\caption{Hubble parameter as a function of redshift for the full UNION2 sample.  Distances in the UNION2 catalog are fit with a cosmological model with $\Omega_m = 0.27$ and $\Omega_{\Lambda} = 0.73$ and are nominally on a scale with H$_0 = 100$~\kmsMpc.  The heavy circles with error bars are averaged values of the Hubble parameter in 0.1 bins in log~$z$ (excluding Hubble parameter values deviating more than a factor 2 from 100).  The averaged values tend to lie slightly below the line at H$_0 = 100$.}
\label{hubpar100}
\end{figure}

We turn attention now to the full UNION2 sample of SNIa and to a best fit for the Hubble Constant over the wide range of redshifts covered by that sample.  Figure~\ref{hub_union} shows the preferred fit of a cosmological model by the UNION2 collaboration \citep{2010ApJ...716..712A} in a Hubble diagram with distances on a relative scale.  We show the same UNION2 data again in Figure~\ref{hubpar100} except we replace distance on the ordinate with the Hubble parameter calculated in accordance with Eq.~1.  This display expands the view of the dispersion around the model fit and the binned data gives clearer insight into systematic deviations.  It is seen that the binned data lies slightly below the nominal model value of H$_0 = 100$~\kmsMpc\ at most redshifts.  

\begin{figure}[!]
\centering
\includegraphics[scale=0.43]{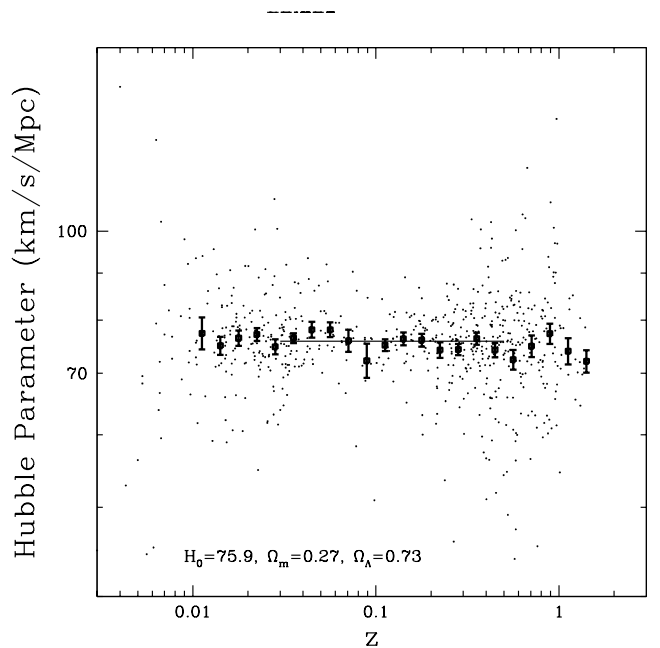}
\caption{Hubble parameter as a function of redshift for the full Union2 sample with zero point shifted to the CF2 scale in accordance with the fit to the 5 source sample of SNIa shown in Fig.~\ref{hub_5_30k}.    Over the range $0.03 < z < 0.5$ the UNION2 sample is fit by H$_0 = 75.9$~\kmsMpc.}
\label{hubpar75}
\end{figure}

The discussion that lead to Figure~\ref{hub_5_30k} provides a robust zero point link to the UNION2 sample.  The shift to this zero point (-0.111 in the logarithm of the Hubble parameter in accordance with the relationship $\mu_{UNION2} = \mu_{CF2} - 0.557$) results in the plot shown in Figure~\ref{hubpar75}.

A least squares fit to log H$_0$ over the full redshift range $z > 0.01$ (629 cases after $3\sigma$ clip) gives the result H$_0 = 75.9$~\kmsMpc.  There is a hint that there is a drop in the Hubble parameter at H$_0 \sim 0.08$.  The best fit over the range $0.03 < z < 0.08$ (97 cases after $3\sigma$ clip) gives H$_0 = 76.6$ while the best fit over the range $0.08 < z < 0.5$ (256 cases, $3\sigma$ clip) gives H$_0 = 75.7$.  These fits exclude $z < 0.03$ where peculiar velocities might be important and $z > 0.5$ where cosmological corrections and errors are important.  The difference in H$_0$ between the nearer and farther range is $0.9 \pm 0.7$, not a significant effect.    We checked for a possible dip in the Hubble parameter near $0.08 < z < 0.1$ in two independent compilations.  In the case of the Carnegie Supernova Project compilation \citep{2009ApJ...704.1036F, 2010AJ....139..120F} there is a step downward at $z \sim 0.1$ of $\Delta {\rm H}_0 = 2.4 \pm 1.4$, again not statistically significant but consistent with UNION2.  However in the \citet{2009ApJ...700.1097H} SALT2 compilation there is essentially no difference in H$_0$ between low and high redshift cuts.   We conclude that the best fit with the UNION2 compilation is obtained by averaging over the range $0.03 < z < 0.5$, with 347 cases ($3\sigma$ clip) giving H$_0 = 75.9 \pm 0.3$ (standard deviation of the mean) with rms scatter $\pm 6.6$~\kmsMpc.

\section{Discussion and Summary}

The value of H$_0$ derived from our calibration is somewhat higher than most recent literature values although within the range of reasonable uncertainties as we will review.  For example \citet{2011ApJ...730..119R} consider three alternative zero point calibrations for the Cepheid Period-Luminosity relation which they link to the SNIa scale and arrive at the preference H$_0 = 73.8 \pm 2.4$~\kmsMpc\ (random + systematic).  Alternatively,  \citet{2006ApJ...653..861M} used a methodology with SFI++ and a calibration based on Cepheid distances with strong parallels with the present work and found H$_0 = 74 \pm 2$ (random) $\pm 6$ (systematic).  In a review, \citet{2010ARA&A..48..673F} advocate H$_0 = 73 \pm 2$ (random) $\pm 4$ (systematic).  A 7 year WMAP analysis   \citep{2011ApJS..192...18K} prefers, but does not directly measure, H$_0 = 71.0 \pm 2.5$~\kmsMpc.

It is not new that we find a slightly higher value for H$_0$.  In our recent re-calibration of the TF relation (Tully \& Courtois 2012) we tentatively found H$_0 = 75$~\kmsMpc\ from distances to 13 clusters, slightly lower than the value found from a calibration a decade earlier  \citep{2000ApJ...533..744T} of H$_0 = 77$~\kmsMpc, mostly, because of a small Malmquist (selection) bias correction now being applied and, slightly, because of the availability of more zero point calibrators.
\citet{2010ARA&A..48..673F}, \citet{2011ApJ...730..119R}, and \citet{2011AJ....142..192F} have extensive discussions of errors.  They argue that uncertainties are presently at the level of 3\% and that 2\% is within reach.  Perhaps so, but we think that there are concerns that are not included in their error budgets.  For one, the scatter in Figure~\ref{sn_tf} is disconcertingly above the expectation of 0.45 if SNIa measure distances with a characteristic uncertainty of 10\%.  In Figure~\ref{dm_gp}, the scatter in the zero point between the separate group or Cepheid based components is 4\%.  This number provides an estimate of systematics.

Our other concern is with the possible influence of peculiar velocities.  \citet{2011ApJ...730..119R} consider the uncertainty here is 0.5\%.  \citet{2011AJ....142..192F} do not budget this effect.  It is possible the tentative dip in the Hubble parameter at $z \sim 0.08$ is a manifestation of very large scale flows.  Flows on the scale of 20,000~\kms\ have been reported by \citet{2011MNRAS.414..264C} and \citet{2011arXiv1111.0631T} based on analyses of two of the SNIa samples used here \citep{2010ApJ...716..712A, 2009ApJ...700.1097H}.  We partially discount the effect of any such flow by extending to $z = 0.5$ with our fit for H$_0$ so a systematic should be 1\% or less over such a great range.

If otherwise we accept the error budget given by \citet{2011ApJ...730..119R} amounting to 3\% then our total estimated error including systematics amounts to 5\%.  We conclude with our best estimate for the Hubble Constant of H$_0 = 75.9 \pm 3.8$~\kmsMpc.

\bigskip\bigskip\bigskip\noindent
Individuals that have helped with the collection and analysis of contributing data include Austin Barnes, Nicolas Bonhomme, Rick Fisher, Philippe H\'eraudeau, Dmitry Makarov, Luca Rizzi, and Max Zavodny.  Material for the Fundamental Plane comparison that supplemented published information was supplied by John Blakeslee and Mike Hudson.  The component of our HI profile information that is new comes from observations in the course of the Cosmic Flows Large Program with the NRAO Green Bank Telescope augmented by observations with Arecibo and Parkes telescopes.  Support has been provided by the US National Science Foundation with award AST-0908846.

\bibliography{paper}

\begin{thebibliography}{42}
\expandafter\ifx\csname natexlab\endcsname\relax\def\natexlab#1{#1}\fi

\bibitem[{{Abell} {et~al.}(1989){Abell}, {Corwin}, \&
  {Olowin}}]{1989ApJS...70....1A}
{Abell}, G.~O., {Corwin}, Jr., H.~G., \& {Olowin}, R.~P. 1989, \apjs, 70, 1

\bibitem[{{Amanullah} {et~al.}(2010){Amanullah}, {Lidman}, {Rubin}, {Aldering},
  {Astier}, {Barbary}, {Burns}, {Conley}, {Dawson}, {Deustua}, {Doi}, {Fabbro},
  {Faccioli}, {Fakhouri}, {Folatelli}, {Fruchter}, {Furusawa}, {Garavini},
  {Goldhaber}, {Goobar}, {Groom}, {Hook}, {Howell}, {Kashikawa}, {Kim}, {Knop},
  {Kowalski}, {Linder}, {Meyers}, {Morokuma}, {Nobili}, {Nordin}, {Nugent},
  {{\"O}stman}, {Pain}, {Panagia}, {Perlmutter}, {Raux}, {Ruiz-Lapuente},
  {Spadafora}, {Strovink}, {Suzuki}, {Wang}, {Wood-Vasey}, {Yasuda}, \&
  {Supernova Cosmology Project}}]{2010ApJ...716..712A}
{Amanullah}, R., {Lidman}, C., {Rubin}, D., {Aldering}, G., {Astier}, P.,
  {Barbary}, K., {Burns}, M.~S., {Conley}, A., {Dawson}, K.~S., {Deustua},
  S.~E., {Doi}, M., {Fabbro}, S., {Faccioli}, L., {Fakhouri}, H.~K.,
  {Folatelli}, G., {Fruchter}, A.~S., {Furusawa}, H., {Garavini}, G.,
  {Goldhaber}, G., {Goobar}, A., {Groom}, D.~E., {Hook}, I., {Howell}, D.~A.,
  {Kashikawa}, N., {Kim}, A.~G., {Knop}, R.~A., {Kowalski}, M., {Linder}, E.,
  {Meyers}, J., {Morokuma}, T., {Nobili}, S., {Nordin}, J., {Nugent}, P.~E.,
  {{\"O}stman}, L., {Pain}, R., {Panagia}, N., {Perlmutter}, S., {Raux}, J.,
  {Ruiz-Lapuente}, P., {Spadafora}, A.~L., {Strovink}, M., {Suzuki}, N.,
  {Wang}, L., {Wood-Vasey}, W.~M., {Yasuda}, N., \& {Supernova Cosmology
  Project}, T. 2010, \apj, 716, 712

\bibitem[{{Bernardi} {et~al.}(2002{\natexlab{a}}){Bernardi}, {Alonso}, {da
  Costa}, {Willmer}, {Wegner}, {Pellegrini}, {Rit{\'e}}, \&
  {Maia}}]{2002AJ....123.2990B}
{Bernardi}, M., {Alonso}, M.~V., {da Costa}, L.~N., {Willmer}, C.~N.~A.,
  {Wegner}, G., {Pellegrini}, P.~S., {Rit{\'e}}, C., \& {Maia}, M.~A.~G.
  2002{\natexlab{a}}, \aj, 123, 2990

\bibitem[{{Bernardi} {et~al.}(2002{\natexlab{b}}){Bernardi}, {Alonso}, {da
  Costa}, {Willmer}, {Wegner}, {Pellegrini}, {Rit{\'e}}, \&
  {Maia}}]{2002AJ....123.2159B}
---. 2002{\natexlab{b}}, \aj, 123, 2159

\bibitem[{{Blakeslee} {et~al.}(2010){Blakeslee}, {Cantiello}, {Mei},
  {C{\^o}t{\'e}}, {Barber DeGraaff}, {Ferrarese}, {Jord{\'a}n}, {Peng},
  {Tonry}, \& {Worthey}}]{2010ApJ...724..657B}
{Blakeslee}, J.~P., {Cantiello}, M., {Mei}, S., {C{\^o}t{\'e}}, P., {Barber
  DeGraaff}, R., {Ferrarese}, L., {Jord{\'a}n}, A., {Peng}, E.~W., {Tonry},
  J.~L., \& {Worthey}, G. 2010, \apj, 724, 657

\bibitem[{{Blakeslee} {et~al.}(2009){Blakeslee}, {Jord{\'a}n}, {Mei},
  {C{\^o}t{\'e}}, {Ferrarese}, {Infante}, {Peng}, {Tonry}, \&
  {West}}]{2009ApJ...694..556B}
{Blakeslee}, J.~P., {Jord{\'a}n}, A., {Mei}, S., {C{\^o}t{\'e}}, P.,
  {Ferrarese}, L., {Infante}, L., {Peng}, E.~W., {Tonry}, J.~L., \& {West},
  M.~J. 2009, \apj, 694, 556

\bibitem[{{Blakeslee} {et~al.}(2001){Blakeslee}, {Lucey}, {Barris}, {Hudson},
  \& {Tonry}}]{2001MNRAS.327.1004B}
{Blakeslee}, J.~P., {Lucey}, J.~R., {Barris}, B.~J., {Hudson}, M.~J., \&
  {Tonry}, J.~L. 2001, \mnras, 327, 1004

\bibitem[{{Catinella} {et~al.}(2007){Catinella}, {Haynes}, \&
  {Giovanelli}}]{2007AJ....134..334C}
{Catinella}, B., {Haynes}, M.~P., \& {Giovanelli}, R. 2007, \aj, 134, 334

\bibitem[{{Colin} {et~al.}(2011){Colin}, {Mohayaee}, {Sarkar}, \&
  {Shafieloo}}]{2011MNRAS.414..264C}
{Colin}, J., {Mohayaee}, R., {Sarkar}, S., \& {Shafieloo}, A. 2011, \mnras,
  414, 264

\bibitem[{{Colless} {et~al.}(2001){Colless}, {Saglia}, {Burstein}, {Davies},
  {McMahan}, \& {Wegner}}]{2001MNRAS.321..277C}
{Colless}, M., {Saglia}, R.~P., {Burstein}, D., {Davies}, R.~L., {McMahan},
  R.~K., \& {Wegner}, G. 2001, \mnras, 321, 277

\bibitem[{{Courtois} {et~al.}(2009){Courtois}, {Tully}, {Fisher}, {Bonhomme},
  {Zavodny}, \& {Barnes}}]{2009AJ....138.1938C}
{Courtois}, H.~M., {Tully}, R.~B., {Fisher}, J.~R., {Bonhomme}, N., {Zavodny},
  M., \& {Barnes}, A. 2009, \aj, 138, 1938

\bibitem[{{Courtois} {et~al.}(2011{\natexlab{a}}){Courtois}, {Tully}, \&
  {H{\'e}raudeau}}]{2011MNRAS.tmp..796C}
{Courtois}, H.~M., {Tully}, R.~B., \& {H{\'e}raudeau}, P. 2011{\natexlab{a}},
  \mnras, 796

\bibitem[{{Courtois} {et~al.}(2011{\natexlab{b}}){Courtois}, {Tully},
  {Makarov}, {Mitronova}, {Koribalski}, {Karachentsev}, \&
  {Fisher}}]{2011MNRAS.414.2005C}
{Courtois}, H.~M., {Tully}, R.~B., {Makarov}, D.~I., {Mitronova}, S.,
  {Koribalski}, B., {Karachentsev}, I.~D., \& {Fisher}, J.~R.
  2011{\natexlab{b}}, \mnras, 414, 2005

\bibitem[{{Dale} {et~al.}(1999){Dale}, {Giovanelli}, {Haynes}, {Campusano}, \&
  {Hardy}}]{1999AJ....118.1489D}
{Dale}, D.~A., {Giovanelli}, R., {Haynes}, M.~P., {Campusano}, L.~E., \&
  {Hardy}, E. 1999, \aj, 118, 1489

\bibitem[{{Folatelli} {et~al.}(2010){Folatelli}, {Phillips}, {Burns},
  {Contreras}, {Hamuy}, {Freedman}, {Persson}, {Stritzinger}, {Suntzeff},
  {Krisciunas}, {Boldt}, {Gonz{\'a}lez}, {Krzeminski}, {Morrell}, {Roth},
  {Salgado}, {Madore}, {Murphy}, {Wyatt}, {Li}, {Filippenko}, \&
  {Miller}}]{2010AJ....139..120F}
{Folatelli}, G., {Phillips}, M.~M., {Burns}, C.~R., {Contreras}, C., {Hamuy},
  M., {Freedman}, W.~L., {Persson}, S.~E., {Stritzinger}, M., {Suntzeff},
  N.~B., {Krisciunas}, K., {Boldt}, L., {Gonz{\'a}lez}, S., {Krzeminski}, W.,
  {Morrell}, N., {Roth}, M., {Salgado}, F., {Madore}, B.~F., {Murphy}, D.,
  {Wyatt}, P., {Li}, W., {Filippenko}, A.~V., \& {Miller}, N. 2010, \aj, 139,
  120

\bibitem[{{Freedman} {et~al.}(2009){Freedman}, {Burns}, {Phillips}, {Wyatt},
  {Persson}, {Madore}, {Contreras}, {Folatelli}, {Gonzalez}, {Hamuy}, {Hsiao},
  {Kelson}, {Morrell}, {Murphy}, {Roth}, {Stritzinger}, {Sturch}, {Suntzeff},
  {Astier}, {Balland}, {Bassett}, {Boldt}, {Carlberg}, {Conley}, {Frieman},
  {Garnavich}, {Guy}, {Hardin}, {Howell}, {Kessler}, {Lampeitl}, {Marriner},
  {Pain}, {Perrett}, {Regnault}, {Riess}, {Sako}, {Schneider}, {Sullivan}, \&
  {Wood-Vasey}}]{2009ApJ...704.1036F}
{Freedman}, W.~L., {Burns}, C.~R., {Phillips}, M.~M., {Wyatt}, P., {Persson},
  S.~E., {Madore}, B.~F., {Contreras}, C., {Folatelli}, G., {Gonzalez}, E.~S.,
  {Hamuy}, M., {Hsiao}, E., {Kelson}, D.~D., {Morrell}, N., {Murphy}, D.~C.,
  {Roth}, M., {Stritzinger}, M., {Sturch}, L., {Suntzeff}, N.~B., {Astier}, P.,
  {Balland}, C., {Bassett}, B., {Boldt}, L., {Carlberg}, R.~G., {Conley},
  A.~J., {Frieman}, J.~A., {Garnavich}, P.~M., {Guy}, J., {Hardin}, D.,
  {Howell}, D.~A., {Kessler}, R., {Lampeitl}, H., {Marriner}, J., {Pain}, R.,
  {Perrett}, K., {Regnault}, N., {Riess}, A.~G., {Sako}, M., {Schneider},
  D.~P., {Sullivan}, M., \& {Wood-Vasey}, M. 2009, \apj, 704, 1036

\bibitem[{{Freedman} \& {Madore}(2010)}]{2010ARA&A..48..673F}
{Freedman}, W.~L. \& {Madore}, B.~F. 2010, \araa, 48, 673

\bibitem[{{Freedman} {et~al.}(2001){Freedman}, {Madore}, {Gibson}, {Ferrarese},
  {Kelson}, {Sakai}, {Mould}, {Kennicutt}, {Ford}, {Graham}, {Huchra},
  {Hughes}, {Illingworth}, {Macri}, \& {Stetson}}]{2001ApJ...553...47F}
{Freedman}, W.~L., {Madore}, B.~F., {Gibson}, B.~K., {Ferrarese}, L., {Kelson},
  D.~D., {Sakai}, S., {Mould}, J.~R., {Kennicutt}, Jr., R.~C., {Ford}, H.~C.,
  {Graham}, J.~A., {Huchra}, J.~P., {Hughes}, S.~M.~G., {Illingworth}, G.~D.,
  {Macri}, L.~M., \& {Stetson}, P.~B. 2001, \apj, 553, 47

\bibitem[{{Freedman} {et~al.}(2011){Freedman}, {Madore}, {Scowcroft}, {Monson},
  {Persson}, {Seibert}, {Rigby}, {Sturch}, \& {Stetson}}]{2011AJ....142..192F}
{Freedman}, W.~L., {Madore}, B.~F., {Scowcroft}, V., {Monson}, A., {Persson},
  S.~E., {Seibert}, M., {Rigby}, J.~R., {Sturch}, L., \& {Stetson}, P. 2011,
  \aj, 142, 192

\bibitem[{{Giovanelli} {et~al.}(1997){Giovanelli}, {Haynes}, {Herter}, {Vogt},
  {da Costa}, {Freudling}, {Salzer}, \& {Wegner}}]{1997AJ....113...53G}
{Giovanelli}, R., {Haynes}, M.~P., {Herter}, T., {Vogt}, N.~P., {da Costa},
  L.~N., {Freudling}, W., {Salzer}, J.~J., \& {Wegner}, G. 1997, \aj, 113, 53

\bibitem[{{Herrnstein} {et~al.}(1999){Herrnstein}, {Moran}, {Greenhill},
  {Diamond}, {Inoue}, {Nakai}, {Miyoshi}, {Henkel}, \&
  {Riess}}]{1999Natur.400..539H}
{Herrnstein}, J.~R., {Moran}, J.~M., {Greenhill}, L.~J., {Diamond}, P.~J.,
  {Inoue}, M., {Nakai}, N., {Miyoshi}, M., {Henkel}, C., \& {Riess}, A. 1999,
  \nat, 400, 539

\bibitem[{{Hicken} {et~al.}(2009){Hicken}, {Wood-Vasey}, {Blondin}, {Challis},
  {Jha}, {Kelly}, {Rest}, \& {Kirshner}}]{2009ApJ...700.1097H}
{Hicken}, M., {Wood-Vasey}, W.~M., {Blondin}, S., {Challis}, P., {Jha}, S.,
  {Kelly}, P.~L., {Rest}, A., \& {Kirshner}, R.~P. 2009, \apj, 700, 1097

\bibitem[{{Hudson} {et~al.}(2001){Hudson}, {Lucey}, {Smith}, {Schlegel}, \&
  {Davies}}]{2001MNRAS.327..265H}
{Hudson}, M.~J., {Lucey}, J.~R., {Smith}, R.~J., {Schlegel}, D.~J., \&
  {Davies}, R.~L. 2001, \mnras, 327, 265

\bibitem[{{Jha} {et~al.}(2007){Jha}, {Riess}, \&
  {Kirshner}}]{2007ApJ...659..122J}
{Jha}, S., {Riess}, A.~G., \& {Kirshner}, R.~P. 2007, \apj, 659, 122

\bibitem[{{Komatsu} {et~al.}(2011){Komatsu}, {Smith}, {Dunkley}, {Bennett},
  {Gold}, {Hinshaw}, {Jarosik}, {Larson}, {Nolta}, {Page}, {Spergel},
  {Halpern}, {Hill}, {Kogut}, {Limon}, {Meyer}, {Odegard}, {Tucker}, {Weiland},
  {Wollack}, \& {Wright}}]{2011ApJS..192...18K}
{Komatsu}, E., {Smith}, K.~M., {Dunkley}, J., {Bennett}, C.~L., {Gold}, B.,
  {Hinshaw}, G., {Jarosik}, N., {Larson}, D., {Nolta}, M.~R., {Page}, L.,
  {Spergel}, D.~N., {Halpern}, M., {Hill}, R.~S., {Kogut}, A., {Limon}, M.,
  {Meyer}, S.~S., {Odegard}, N., {Tucker}, G.~S., {Weiland}, J.~L., {Wollack},
  E., \& {Wright}, E.~L. 2011, \apjs, 192, 18

\bibitem[{{Masters} {et~al.}(2006){Masters}, {Springob}, {Haynes}, \&
  {Giovanelli}}]{2006ApJ...653..861M}
{Masters}, K.~L., {Springob}, C.~M., {Haynes}, M.~P., \& {Giovanelli}, R. 2006,
  \apj, 653, 861

\bibitem[{{Perlmutter} {et~al.}(1999){Perlmutter}, {Aldering}, {Goldhaber},
  {Knop}, {Nugent}, {Castro}, {Deustua}, {Fabbro}, {Goobar}, {Groom}, {Hook},
  {Kim}, {Kim}, {Lee}, {Nunes}, {Pain}, {Pennypacker}, {Quimby}, {Lidman},
  {Ellis}, {Irwin}, {McMahon}, {Ruiz-Lapuente}, {Walton}, {Schaefer}, {Boyle},
  {Filippenko}, {Matheson}, {Fruchter}, {Panagia}, {Newberg}, {Couch}, \& {The
  Supernova Cosmology Project}}]{1999ApJ...517..565P}
{Perlmutter}, S., {Aldering}, G., {Goldhaber}, G., {Knop}, R.~A., {Nugent}, P.,
  {Castro}, P.~G., {Deustua}, S., {Fabbro}, S., {Goobar}, A., {Groom}, D.~E.,
  {Hook}, I.~M., {Kim}, A.~G., {Kim}, M.~Y., {Lee}, J.~C., {Nunes}, N.~J.,
  {Pain}, R., {Pennypacker}, C.~R., {Quimby}, R., {Lidman}, C., {Ellis}, R.~S.,
  {Irwin}, M., {McMahon}, R.~G., {Ruiz-Lapuente}, P., {Walton}, N., {Schaefer},
  B., {Boyle}, B.~J., {Filippenko}, A.~V., {Matheson}, T., {Fruchter}, A.~S.,
  {Panagia}, N., {Newberg}, H.~J.~M., {Couch}, W.~J., \& {The Supernova
  Cosmology Project}. 1999, \apj, 517, 565

\bibitem[{{Prieto} {et~al.}(2006){Prieto}, {Rest}, \&
  {Suntzeff}}]{2006ApJ...647..501P}
{Prieto}, J.~L., {Rest}, A., \& {Suntzeff}, N.~B. 2006, \apj, 647, 501

\bibitem[{{Riess} {et~al.}(1998){Riess}, {Filippenko}, {Challis},
  {Clocchiatti}, {Diercks}, {Garnavich}, {Gilliland}, {Hogan}, {Jha},
  {Kirshner}, {Leibundgut}, {Phillips}, {Reiss}, {Schmidt}, {Schommer},
  {Smith}, {Spyromilio}, {Stubbs}, {Suntzeff}, \&
  {Tonry}}]{1998AJ....116.1009R}
{Riess}, A.~G., {Filippenko}, A.~V., {Challis}, P., {Clocchiatti}, A.,
  {Diercks}, A., {Garnavich}, P.~M., {Gilliland}, R.~L., {Hogan}, C.~J., {Jha},
  S., {Kirshner}, R.~P., {Leibundgut}, B., {Phillips}, M.~M., {Reiss}, D.,
  {Schmidt}, B.~P., {Schommer}, R.~A., {Smith}, R.~C., {Spyromilio}, J.,
  {Stubbs}, C., {Suntzeff}, N.~B., \& {Tonry}, J. 1998, \aj, 116, 1009

\bibitem[{{Riess} {et~al.}(2011){Riess}, {Macri}, {Casertano}, {Lampeitl},
  {Ferguson}, {Filippenko}, {Jha}, {Li}, \& {Chornock}}]{2011ApJ...730..119R}
{Riess}, A.~G., {Macri}, L., {Casertano}, S., {Lampeitl}, H., {Ferguson},
  H.~C., {Filippenko}, A.~V., {Jha}, S.~W., {Li}, W., \& {Chornock}, R. 2011,
  \apj, 730, 119

\bibitem[{{Riess} {et~al.}(2009){Riess}, {Macri}, {Li}, {Lampeitl},
  {Casertano}, {Ferguson}, {Filippenko}, {Jha}, {Chornock}, {Greenhill},
  {Mutchler}, {Ganeshalingham}, \& {Hicken}}]{2009ApJS..183..109R}
{Riess}, A.~G., {Macri}, L., {Li}, W., {Lampeitl}, H., {Casertano}, S.,
  {Ferguson}, H.~C., {Filippenko}, A.~V., {Jha}, S.~W., {Chornock}, R.,
  {Greenhill}, L., {Mutchler}, M., {Ganeshalingham}, M., \& {Hicken}, M. 2009,
  \apjs, 183, 109

\bibitem[{{Rizzi} {et~al.}(2007){Rizzi}, {Tully}, {Makarov}, {Makarova},
  {Dolphin}, {Sakai}, \& {Shaya}}]{2007ApJ...661..815R}
{Rizzi}, L., {Tully}, R.~B., {Makarov}, D., {Makarova}, L., {Dolphin}, A.~E.,
  {Sakai}, S., \& {Shaya}, E.~J. 2007, \apj, 661, 815

\bibitem[{{Sakai} {et~al.}(2004){Sakai}, {Ferrarese}, {Kennicutt}, \&
  {Saha}}]{2004ApJ...608...42S}
{Sakai}, S., {Ferrarese}, L., {Kennicutt}, Jr., R.~C., \& {Saha}, A. 2004,
  \apj, 608, 42

\bibitem[{{Springob} {et~al.}(2007){Springob}, {Masters}, {Haynes},
  {Giovanelli}, \& {Marinoni}}]{2007ApJS..172..599S}
{Springob}, C.~M., {Masters}, K.~L., {Haynes}, M.~P., {Giovanelli}, R., \&
  {Marinoni}, C. 2007, \apjs, 172, 599

\bibitem[{{Tonry} {et~al.}(2001){Tonry}, {Dressler}, {Blakeslee}, {Ajhar},
  {Fletcher}, {Luppino}, {Metzger}, \& {Moore}}]{2001ApJ...546..681T}
{Tonry}, J.~L., {Dressler}, A., {Blakeslee}, J.~P., {Ajhar}, E.~A., {Fletcher},
  A.~B., {Luppino}, G.~A., {Metzger}, M.~R., \& {Moore}, C.~B. 2001, \apj, 546,
  681

\bibitem[{{Tonry} {et~al.}(2003){Tonry}, {Schmidt}, {Barris}, {Candia},
  {Challis}, {Clocchiatti}, {Coil}, {Filippenko}, {Garnavich}, {Hogan},
  {Holland}, {Jha}, {Kirshner}, {Krisciunas}, {Leibundgut}, {Li}, {Matheson},
  {Phillips}, {Riess}, {Schommer}, {Smith}, {Sollerman}, {Spyromilio},
  {Stubbs}, \& {Suntzeff}}]{2003ApJ...594....1T}
{Tonry}, J.~L., {Schmidt}, B.~P., {Barris}, B., {Candia}, P., {Challis}, P.,
  {Clocchiatti}, A., {Coil}, A.~L., {Filippenko}, A.~V., {Garnavich}, P.,
  {Hogan}, C., {Holland}, S.~T., {Jha}, S., {Kirshner}, R.~P., {Krisciunas},
  K., {Leibundgut}, B., {Li}, W., {Matheson}, T., {Phillips}, M.~M., {Riess},
  A.~G., {Schommer}, R., {Smith}, R.~C., {Sollerman}, J., {Spyromilio}, J.,
  {Stubbs}, C.~W., \& {Suntzeff}, N.~B. 2003, \apj, 594, 1

\bibitem[{{Tully}(1987)}]{1987ApJ...321..280T}
{Tully}, R.~B. 1987, \apj, 321, 280

\bibitem[{{Tully}(1988)}]{1988ngc..book.....T}
---. 1988, {Nearby galaxies catalog} (Cambridge and New York, Cambridge
  University Press, 1988, 221 p.)

\bibitem[{{Tully} \& {Courtois}(2012)}]{2012arXiv1202.3191T}
{Tully}, R.~B. \& {Courtois}, H.~M. 2012, ArXiv e-prints

\bibitem[{{Tully} \& {Fisher}(1977)}]{1977A&A....54..661T}
{Tully}, R.~B. \& {Fisher}, J.~R. 1977, \aap, 54, 661

\bibitem[{{Tully} \& {Pierce}(2000)}]{2000ApJ...533..744T}
{Tully}, R.~B. \& {Pierce}, M.~J. 2000, \apj, 533, 744

\bibitem[{{Turnbull} {et~al.}(2011){Turnbull}, {Hudson}, {Feldman}, {Hicken},
  {Kirshner}, \& {Watkins}}]{2011arXiv1111.0631T}
{Turnbull}, S.~J., {Hudson}, M.~J., {Feldman}, H.~A., {Hicken}, M., {Kirshner},
  R.~P., \& {Watkins}, R. 2011, ArXiv e-prints

\end{thebibliography}
\bibliographystyle{apj}

\begin{deluxetable}{rllrcccrcccc}

\tablenum{2}
\tablecaption{Distances to Galaxies that have Hosted SNIa}
\label{tbl:sn_tf}
\tablewidth{0in}

\tablehead{\colhead{PGC} & \colhead{Name} & \colhead{SNIa} & \colhead{$V_{CMB}$} & \colhead{Incl.} & \colhead{$W_{mx}$} & \colhead{$W_{mx}^i$} & \colhead{$I_T^{bik}$} & \colhead{$M_I^{bik}$} & \colhead{$\mu_{TF}$} & \colhead{So. SN} & \colhead{$\mu_{SN}$}}
\startdata
    250 & UGC00014   & 2006sr &   6916 &  56 &   331 &   398 &  12.01 &  -22.27 &  34.33 & u  h  &  34.34 \\
    415 & UGC00040   & 2003it &   7198 &  55 &   372 &   451 &  12.42 &  -22.75 &  35.26 & u  h  &  34.44 \\
    963 & UGC00139   & 1998dk &   3614 &  67 &   287 &   311 &  12.51 &  -21.33 &  33.88 & upj   &  32.75 \\
   1288 & PGC001288  & 1999cw &   3370 &  66 &   282 &   308 &  12.54 &  -21.29 &  33.86 &  pj   &  32.53 \\
   3773 & UGC00646   & 1998ef &   5011 &  71 &   367 &   389 &  11.76 &  -22.18 &  33.99 & upjh  &  33.25 \\
   4915 & NGC0477    & 2002jy &   5600 &  61 &   322 &   369 &  11.76 &  -21.98 &  33.78 & u  h  &  34.31 \\
   5341 & PGC005341  & 1998dm &   1663 &  90 &   236 &   236 &  11.54 &  -20.27 &  31.81 & u j   &  32.24 \\
   6624 & NGC0673    & 1996bo &   4898 &  45 &   315 &   445 &  11.25 &  -22.69 &  33.99 & upj   &  33.23 \\
   9560 & NGC0958    & 2005A  &   5501 &  74 &   561 &   584 &  10.30 &  -23.74 &  34.09 &     f &  33.82 \\
   9618 & UGC01993   & 1999gp &   7812 &  85 &   483 &   485 &  11.84 &  -23.03 &  34.94 & upjh  &  34.63 \\
  10448 & IC1844     & 1995ak &   6588 &  72 &   294 &   309 &  11.87 &  -21.30 &  33.20 & upjh  &  33.87 \\
  11606 & ESO300-009 & 1992bc &   5918 &  73 &   309 &   323 &  13.55 &  -21.47 &  35.10 & upjh  &  33.91 \\
  13727 & NGC1448    & 2001el &   1092 &  90 &   386 &   386 &   8.94 &  -22.15 &  31.10 &   j   &  30.61 \\
  17509 & UGC03329   & 1999ek &   5277 &  67 &   484 &   525 &  10.50 &  -23.33 &  33.87 & upjh  &  33.53 \\
  18089 & UGC03375   & 2001gc &   5792 &  66 &   490 &   535 &  10.63 &  -23.40 &  34.07 & u     &  33.48 \\
  18373 & PGC018373  & 2003kf &   2295 &  90 &   234 &   234 &  11.55 &  -20.24 &  31.79 & u     &  31.85 \\
  18747 & UGC03432   & 1996bv &   5015 &  80 &   285 &   289 &  12.92 &  -21.05 &  34.01 & upjh  &  33.32 \\
  19788 & UGC03576   & 1998ec &   6013 &  65 &   356 &   393 &  12.00 &  -22.22 &  34.27 & upj   &  34.07 \\
  20513 & UGC03770   & 2000fa &   6525 &  58 &   314 &   371 &  12.54 &  -22.00 &  34.61 & upjh  &  34.21 \\
  21020 & UGC03845   & 1997do &   3136 &  49 &   195 &   257 &  12.32 &  -20.60 &  32.94 & upjh  &  32.67 \\
  26512 & NGC2841    & 1999by &    804 &  66 &   592 &   650 &   7.52 &  -24.14 &  31.67 &   j   &  30.23 \\
  28357 & NGC3021    & 1995al &   1797 &  57 &   254 &   303 &  10.92 &  -21.23 &  32.15 & u j   &  31.73 \\
  31428 & NGC3294    & 1992G  &   1831 &  64 &   386 &   431 &   9.81 &  -22.58 &  32.40 &   j   &  31.65 \\
  32192 & NGC3368    & 1998bu &   1231 &  50 &   328 &   428 &   7.88 &  -22.55 &  30.43 &   j   &  29.35 \\
  32207 & NGC3370    & 1994ae &   1609 &  58 &   264 &   312 &  10.85 &  -21.34 &  32.19 & u j   &  31.58 \\
  34695 & NGC3627    & 1989b  &   1061 &  60 &   333 &   385 &   7.38 &  -22.14 &  29.53 &   j   &  29.11 \\
  35006 & NGC3663    & 2006ax &   5396 &  45 &   314 &   443 &  11.26 &  -22.68 &  33.99 & u  hf &  33.67 \\
  35088 & NGC3672    & 2007bm &   2223 &  71 &   377 &   399 &   9.67 &  -22.28 &  31.95 & u     &  31.62 \\
  36832 & NGC3891    & 2006or &   6573 &  44 &   391 &   561 &  11.39 &  -23.58 &  35.05 & u  h  &  34.25 \\
  41517 & NGC4501    & 1999cl &   2601 &  63 &   507 &   570 &   7.86 &  -23.64 &  31.51 & u j   &  30.35 \\
  41789 & NGC4527    & 1991T  &   2072 &  77 &   352 &   362 &   8.70 &  -21.90 &  30.60 &   j   &  29.84 \\
  41823 & NGC4536    & 1981B  &   2144 &  71 &   322 &   341 &   9.03 &  -21.68 &  30.71 &   j   &  30.18 \\
  42741 & NGC4639    & 1990N  &   1308 &  55 &   274 &   336 &  10.18 &  -21.62 &  31.80 &   j   &  31.07 \\
  43118 & NGC4680    & 1997bp &   2824 &  52 &   186 &   237 &  11.29 &  -20.29 &  31.58 & upj   &  31.97 \\
  43170 & NGC4679    & 2001cz &   4935 &  69 &   399 &   427 &  10.87 &  -22.54 &  33.43 & upjh  &  33.33 \\
  45749 & NGC5005    & 1996ai &   1178 &  63 &   535 &   601 &   8.07 &  -23.85 &  31.92 & u j   &  30.96 \\
  46574 & ESO576-040 & 1997br &   2385 &  85 &   169 &   170 &  12.44 &  -19.01 &  31.45 &   j   &  31.32 \\
  47422 & NGC5185    & 2006br &   7656 &  76 &   587 &   605 &  11.49 &  -23.87 &  35.46 & u     &  34.95 \\
  47514 & PGC047514  & 2007ca &   4517 &  86 &   284 &   285 &  12.62 &  -20.99 &  33.64 & u  h  &  33.78 \\
  50042 & NGC5440    & 1998D  &   3890 &  64 &   610 &   677 &  10.30 &  -24.30 &  34.67 &  pj   &  33.06 \\
  51344 & NGC5584    & 2007af &   1890 &  44 &   186 &   267 &  10.62 &  -20.74 &  31.36 & u     &  31.31 \\
  51549 & IC4423     & 2001ay &   9271 &  61 &   413 &   470 &  13.02 &  -22.91 &  36.06 & u j   &  35.09 \\
  56537 & IC1151     & 1991M  &   2274 &  69 &   226 &   242 &  11.76 &  -20.36 &  32.13 &   j   &  32.58 \\
  57205 & NGC6063    & 1999ac &   2950 &  59 &   265 &   308 &  11.84 &  -21.30 &  33.16 & upj   &  32.41 \\
  59769 & UGC10738   & 2001cp &   6726 &  90 &   585 &   585 &  11.60 &  -23.74 &  35.44 & u  h  &  34.23 \\
  59782 & UGC10743   & 2002er &   2574 &  79 &   202 &   206 &  11.84 &  -19.74 &  31.58 &   j   &  32.10 \\
  64600 & NGC6916    & 2002cd &   2932 &  52 &   340 &   434 &  10.52 &  -22.60 &  33.14 & u     &  32.67 \\
  65375 & NGC6962    & 2002ha &   3936 &  48 &   474 &   633 &  10.18 &  -24.05 &  34.28 & u  h  &  33.10 \\
  66579 & UGC11723   & 2006cm &   4585 &  90 &   426 &   426 &  11.93 &  -22.53 &  34.52 & u     &  33.93 \\
  68455 & IC5179     & 1999ee &   3158 &  63 &   395 &   444 &  10.38 &  -22.69 &  33.09 &  pjh  &  32.59 \\
  69428 & UGC12133   & 1998eg &   7068 &  90 &   443 &   443 &  12.38 &  -22.68 &  35.14 & upjh  &  34.42 \\
  69453 & NGC7329    & 2006bh &   3143 &  53 &   369 &   461 &  10.40 &  -22.83 &  33.26 &     f &  32.61 \\
  70213 & NGC7448    & 1997dt &   1838 &  63 &   281 &   316 &  10.41 &  -21.39 &  31.79 & u j   &  32.49 \\
  71166 & UGC12538   & 2006b  &   4583 &  72 &   278 &   292 &  12.70 &  -21.08 &  33.82 & u     &  33.44 \\
  71534 & NGC7678    & 2002dp &   3143 &  45 &   288 &   407 &  10.10 &  -22.35 &  32.46 & u  h  &  32.68 \\
  72775 & NGC7780    & 2001da &   4853 &  61 &   381 &   437 &  12.16 &  -22.63 &  34.86 & u  h  &  33.44 \\
\enddata
\tablecomments{
(1) Principal Galaxies Catalog number. (2) Common name. (3) Supernova designation. (4) Velocity of host galaxy in CMB frame (\kms). (5) Host galaxy inclination (degrees from face-on). (6) Line width parameter (\kms). (7) Line width parameter projected to edge-on orientation (\kms). (8) $I$ band magnitude adjusted for Galactic and internal absorption and $k$ correction (mag). (9) Absolute $I$ magnitude from line width and TF relation (mag). (10) Distance modulus from TF relation (mag). (11) Sources contributing to SNIa modulus: u=\citet{2010ApJ...716..712A}, p=\citet{2006ApJ...647..501P}, j=\citet{2007ApJ...659..122J}, h=\citet{2009ApJ...700.1097H}, f=\citet{2010AJ....139..120F}, (12) Distance modulus from SNIa light curve, Union2 scale (mag).}

\end{deluxetable}

\begin{deluxetable}{lcccccccccccl}
\tablenum{4}
\tablecaption{Clusters Beyond 3000~\kms}
\label{tbl:dm_gt3k}
\tablewidth{0in}
\tablehead{\colhead{Cluster} & \colhead{\#} & \colhead{$\mu_{other}$} & \colhead{Err} & \colhead{EF} & \colhead{EN} & \colhead{SM} & \colhead{SFI} & \colhead{CF2} & \colhead{\#} & \colhead{$\mu_{SN}$} & \colhead{Err} & \colhead{SNIa Names}}
\startdata
Cen45   & 2 & 33.27 & 0.13 &    &  8 &    &   6 &     & 1 & 33.33 & 0.20 & 2001cz                             \\
J4      & 1 & 35.14 & 0.20 &  2 &    &    &     &     & 1 & 35.65 & 0.20 & 1999ef                             \\
A85     & 1 & 37.37 & 0.20 &  4 &    &    &     &     & 1 & 35.88 & 0.20 & 2003ic                             \\
Pisces  & 7 & 34.00 & 0.05 &  7 & 39 & 46 &  50 &  58 & 5 & 33.44 & 0.09 & 1998ef,1999ej,2000dk,2001en,2006td \\
A194    & 2 & 34.16 & 0.08 &    & 15 & 16 &  23 &     & 1 & 33.59 & 0.20 & 1993ae                             \\
J7      & 1 & 35.76 & 0.20 &  4 &    &    &     &     & 2 & 34.95 & 0.14 & 1999gp,2002hu                      \\
A397    & 2 & 35.41 & 0.09 &  7 &    &    &  14 &     & 1 & 34.90 & 0.20 & 2006os                             \\
Perseus & 2 & 34.02 & 0.11 &    & 26 & 28 &   6 &     & 1 & 33.65 & 0.20 & 2008L                              \\
J28     & 1 & 34.77 & 0.20 &  3 &    &    &     &     & 1 & 34.84 & 0.20 & 2005eq                             \\
P597-1  & 1 & 34.19 & 0.20 &  3 &    &    &     &     & 1 & 32.95 & 0.20 & 2001ep                             \\
A569    & 2 & 34.52 & 0.10 &    &    & 13 &  13 &     & 2 & 33.81 & 0.14 & 2000B,2007au                       \\
Cancer  & 2 & 34.04 & 0.07 &    &    &    &  17 &  11 & 1 & 33.49 & 0.20 & 1999aa                             \\
A999    & 1 & 35.51 & 0.20 &    &    &  5 &     &     & 1 & 35.13 & 0.20 & 2004L                              \\
A1367   & 3 & 34.80 & 0.07 &    &  6 &  8 &  32 &  19 & 1 & 33.75 & 0.20 & 2007ci                             \\
Coma    & 3 & 34.83 & 0.06 & 19 & 80 & 56 &  34 &  23 & 2 & 34.56 & 0.14 & 2006cg,2007bz                      \\
J11     & 1 & 35.30 & 0.20 &  4 &    &    &     &     & 1 & 34.47 & 0.20 & 2007F                              \\
A1736   & 2 & 35.67 & 0.11 &    &    &  4 &  10 &     & 3 & 34.72 & 0.12 & 1991U,1992ag,2007cg                \\
A3558   & 2 & 36.06 & 0.12 &    &    & 28 &   8 &     & 1 & 36.13 & 0.20 & 1993O                              \\
A1983   & 1 & 36.62 & 0.20 &  5 &    &    &     &     & 1 & 35.03 & 0.20 & 2008af                             \\
P445-2  & 1 & 33.77 & 0.20 &  2 &    &    &     &     & 1 & 33.64 & 0.20 & 2007ap                             \\
J20     & 1 & 35.61 & 0.20 &  2 &    &    &     &     & 1 & 34.78 & 0.20 & 2002de                             \\
PavoII  & 2 & 33.71 & 0.10 &    & 12 &  9 &   8 &     & 1 & 33.20 & 0.20 & 2001cn                             \\
A2634   & 3 & 35.39 & 0.07 & 10 & 12 & 32 &  18 &  13 & 1 & 35.17 & 0.20 & 1997dg                             \\
A2666   & 2 & 34.99 & 0.11 &  2 &    &    &   9 &     & 1 & 34.40 & 0.20 & 2007qe                             \\
A4038   & 2 & 35.37 & 0.10 &    & 18 & 18 &   7 &     & 1 & 34.65 & 0.20 & 1993ah                             \\
\enddata
\tablecomments{
Column (1) Cluster name, (2) No. of measures: FP+SFI+CF2 (Pisces is a special case because we consider the Pisces filament as a single unit whereas others break it into pieces - see discussion by  Tully \& Courtois 2012), (3) averaged FP+SFI+CF2 distance modulus, (4) assigned uncertainty, (5) No. EFAR galaxies, (6) No. ENEAR galaxies, (7) No. SMAC galaxies, (8) No. SFI++ galaxies, (9) No. CF2 galaxies, (10) No. of SNIa associated with cluster (Pisces is special case) , (11) averaged SNIa modulus, Union2 scale, (12) assigned error, (13) SNIa names.
}
\end{deluxetable}

\begin{deluxetable}{rccccccccccccccl}
\tablenum{5}
\tablecaption{Groups Within 3000~\kms}
\label{tbl:dm_lt3k}
\tablewidth{0in}
\tablehead{\colhead{ID} & \colhead{NBG} & \colhead{\#} & \colhead{$\mu_{SN}$} & \colhead{Err} & \colhead{\#} & \colhead{$\mu_{SBF}$} & \colhead{\#} & \colhead{$\mu_{Ceph}$} & \colhead{\#} & \colhead{$\mu_{RGB}$} & \colhead{\#} & \colhead{$\mu_{TF}$} & \colhead{$\mu_{av}$} & \colhead{Err} & \colhead{SNIa ID}}
\startdata
   1 & 11 -1 &  4 & 30.27 & 0.10 &  77 & 31.09 &   4 & 30.98 &    &       & 26 & 31.02 & 31.08 & 0.06 & 1990N, 1991bg, 1994D, 1999cl 2006X \\ 
   4 & 11 -4 &  2 & 30.01 & 0.14 &   1 & 31.08 &   2 & 30.88 &    &       &  4 & 30.86 & 30.90 & 0.10 & 1981B, 1991T \\
  51 & 11-22 &  2 & 31.37 & 0.14 &     &       &     &       &    &       &  2 & 31.94 & 31.94 & 0.28 & 1996X, 1997br \\
  65 & 11-27 &  1 & 31.97 & 0.20 &     &       &     &       &    &       &  5 & 32.15 & 32.15 & 0.18 & 1997bp \\
  93 & 11 -0 &  1 & 32.08 & 0.20 &     &       &     &       &    &       &  3 & 32.57 & 32.57 & 0.23 & 2002dj \\
 240 & 14-15 &  1 & 26.53 & 0.20 &     &       &     &       &  1 & 27.77 &    &       & 27.77 & 0.20 & 1986G  \\
 240 & 14-15 &  1 & 26.90 & 0.20 &     &       &     &       &  1 & 27.66 &    &       & 27.66 & 0.20 & 1972E  \\
 266 & 15 -1 &  1 & 29.35 & 0.20 &   6 & 30.24 &   2 & 30.06 &  1 & 30.26 &  1 & 30.38 & 30.21 & 0.07 & 1998bu \\
 267 & 15 -2 &  1 & 29.11 & 0.20 &     &       &   1 & 30.01 &    &       &  2 & 30.01 & 30.01 & 0.16 & 1989B  \\
 293 & 15+10 &  1 & 30.23 & 0.20 &     &       &   1 & 30.74 &    &       &  1 & 31.67 & 30.93 & 0.18 & 1999by \\
 361 & 21 -3 &  1 & 31.58 & 0.20 &   1 & 31.58 &   1 & 32.13 &    &       &  1 & 32.19 & 31.98 & 0.15 & 1994ae \\
 368 & 21 -6 &  1 & 31.15 & 0.20 &   2 & 32.27 &     &       &    &       &  2 & 31.90 & 32.20 & 0.13 & 2002bo \\
 372 & 21-10 &  1 & 31.25 & 0.20 &   1 & 31.76 &     &       &    &       &  1 & 33.08 & 32.03 & 0.18 & 2003cg \\
 378 & 21-12 &  1 & 31.73 & 0.20 &     &       &   1 & 32.27 &    &       &  4 & 31.74 & 32.00 & 0.14 & 1995al \\
 404 & 21 -0 &  1 & 31.65 & 0.20 &     &       &     &       &    &       &  1 & 32.38 & 32.38 & 0.40 & 1992G  \\
 409 & 22 -1 &  1 & 30.93 & 0.20 &   1 & 31.68 &   1 & 31.66 &    &       &  5 & 31.78 & 31.71 & 0.12 & 2007sr \\
 422 & 22 -7 &  1 & 31.62 & 0.20 &     &       &     &       &    &       &  1 & 31.96 & 31.96 & 0.40 & 2007bm \\
 441 & 23 -1 &  1 & 33.33 & 0.20 &   6 & 32.53 &   1 & 32.60 &    &       & 11 & 32.91 & 32.65 & 0.06 & 2001cz \\
 520 & 31-12 &  2 & 31.88 & 0.14 &     &       &     &       &    &       &  1 & 32.18 & 32.18 & 0.40 & 1996Z, 1999gh \\
 555 & 31 -0 &  1 & 31.88 & 0.20 &     &       &     &       &    &       &  1 & 32.58 & 32.58 & 0.40 & 1995D  \\
 627 & 34 +6 &  1 & 31.85 & 0.20 &     &       &     &       &    &       &  1 & 31.78 & 31.78 & 0.40 & 2003kf \\
 671 & 41 +2 &  1 & 31.31 & 0.20 &     &       &   1 & 31.72 &    &       &  1 & 31.36 & 31.65 & 0.18 & 2007af \\
 739 & 42 -7 &  1 & 32.26 & 0.20 &   1 & 32.29 &     &       &    &       &  1 & 32.60 & 32.36 & 0.18 & 2003du \\
 742 & 42 -8 &  1 & 31.52 & 0.20 &   1 & 32.54 &     &       &    &       &  2 & 32.65 & 32.59 & 0.16 & 1996bk \\
 797 & 43 -1 &  1 & 30.59 & 0.20 &     &       &     &       &    &       &  5 & 31.35 & 31.35 & 0.18 & 1996ai \\
 843 & 51 -1 &  2 & 30.70 & 0.14 &  42 & 31.51 &   2 & 31.16 &    &       & 15 & 31.20 & 31.47 & 0.06 & 1980N, 1992A \\
 867 & 51-12 &  1 & 31.80 & 0.20 &     &       &   1 & 32.59 &    &       &  1 & 31.98 & 32.48 & 0.18 & 2002fk \\
 895 & 52 -7 &  1 & 32.24 & 0.20 &   3 & 31.86 &     &       &    &       &  2 & 32.07 & 31.89 & 0.11 & 1998dm \\
 950 & 53 -7 &  1 & 30.61 & 0.20 &   2 & 31.34 &     &       &    &       &  9 & 31.14 & 31.23 & 0.10 & 2001el \\
1167 & 64 -1 &  1 & 32.14 & 0.20 &   1 & 31.89 &     &       &    &       &  3 & 32.05 & 31.96 & 0.15 & 1997dt \\
1179 & 64 -9 &  1 & 31.94 & 0.20 &     &       &     &       &    &       &  1 & 32.58 & 32.58 & 0.40 & 1998dh \\
1232 & 71 -3 &  1 & 32.58 & 0.20 &     &       &     &       &    &       &  1 & 32.12 & 32.12 & 0.40 & 1991M  \\
1249 & 71 -0 &  1 & 32.41 & 0.20 &     &       &     &       &    &       &  1 & 33.15 & 33.15 & 0.40 & 1999ac \\
1319 & 70 -0 &  1 & 32.10 & 0.20 &     &       &     &       &    &       &  1 & 31.58 & 31.58 & 0.40 & 2002er \\
\enddata
\tablecomments{
Column (1) Group ID used in EDD\footnote{http://edd.ifa.hawaii.edu}, (2) Group ID in Nearby Galaxies Catalog \citep{1988ngc..book.....T}, (3) No. SNIa in group, (4) average SNIa distance modulus on Union2 scale, (5) error$=0.2/\sqrt{N}$, (6) No. SBF measures in group, (7) average SBF distance modulus, (8) No. Cepheid measures in group, (9) average Cepheid distance modulus, (10) No. TRGB measures, (11) TRGB distance modulus, (12) No. TF measures in group, (13) average TF distance modulus, (14) Weighted average of SBF, Cepheid, TRGB, and TF distance moduli, (15) assigned error, (16) SNIa name.
}
\end{deluxetable}

\end{document}